\documentclass{ws-mpla}
\usepackage[super]{cite}
\usepackage{graphicx}
\usepackage{amsmath,amssymb}

\def\Journal#1#2#3#4{{#1} {\bf #2}, #3 (#4)}

\def\Sym{\em Symmetry}

\def\NPB{\em Nucl. Phys. B}
\def\PLB{\em Phys. Lett.  B}
\def\PRL{\em Phys. Rev. Lett.}
\def\PRD{\em Phys. Rev. D}

\def\GaC{\em Gravitation and Cosmology}

\def\JETP{\em JETP}
\def\MNRAS{\em Mon. Not. Roy. astr. Soc.}
\def\JETPL{\em JETP Lett.}

\def\IJMPA{\em Int. J. Mod. Phys. A}
\def\IJMPD{\em Int. J. Mod. Phys.  D}
\def\NJP{\em New J. of Phys.}
\def\ARAA{\em Ann. Rev. Astron. Astrophys.}
\def\AIPCP{\em AIP Conf. Proc.}

\def\SJNP{\em Sov.J.Nucl.Phys.}

\def\BWP{\em Bled Workshops in Physics}

\def\({\left(}
\def\){\right)}

\def\beq{\begin{equation}}
\def\eeq{\end{equation}}
\def\bea{\begin{eqnarray}}
\def\eea{\end{eqnarray}}
\begin{document}


\catchline{}{}{}{}{}

\title{Dark matter reflection of particle symmetry} 

\author{\footnotesize Maxim Yu. Khlopov${}^{a,b}$}
\address{${}^a$Laboratory of Astroparticle Physics and Cosmology 10, rue Alice Domon et Léonie Duquet, 75205 Paris Cedex 13, France\\
${}^b$Center for Cosmopartcile Physics “Cosmion” and National Research Nuclear University “MEPHI”
(Moscow State Engineering Physics Institute), Kashirskoe Shosse 31, Moscow 115409, Russia\\
khlopov@apc.univ-paris7.fr}

\maketitle


\begin{abstract}
In the context of the relationship between physics of cosmological dark matter and  symmetry of elementary particles a wide list of dark matter candidates is possible. New symmetries provide stability of different new particles and their combination can lead to a multicomponent dark matter. The pattern of symmetry breaking involves phase transitions in very early Universe, extending the list of candidates by topological defects and even primordial nonlinear structures. 
 
\end{abstract}

\ccode{PACS Nos.: 95.30.Cq; 95.35.+d; 98.80.-k; 12.60.Nz; 11.15.Ex; 11.27.+d; 11.30.Hv; 11.30.Ly}

\section{Introduction}

The structure and interactions of known particles are described on the basis of the principle of gauge symmetry of the Standard model, extending invariance of quantum electrodynamics relative to gauge transformations to symmetry of strong and weak interactions. This approach assumes symmetry between different particles and ascribes their difference to the mechanisms of symmetry breaking. However, the Standard model (SM), successfully describing properties and interactions of the known particles, is not sufficient to provide the basis for the modern inflationary cosmology with baryosynthesis and dark matter/energy, as well as it should be extended to resolve its internal problems like divergence of the mass of the Higgs boson or problem of CP violation in Quantum Chromodynamics (QCD). The possibility to unify strong and electroweak interactions in the framework of Grand Unified Theories (GUT) adds an aesthetical argument to extend the SM. The discovery of nonzero mass of neutrino has already moved physics beyond the SM, in which neutrinos are massless.

 Extensions of the standard model
involve new symmetries and new particle states. Noether's theorem relates the exact particle symmetry to conservation of the respective charge. If the symmetry is strict, the charge is strictly conserved and the lightest particles bearing it are
stable. Born in the early Universe they should be present now around us. Their absence means that they should be elusive, being a form
of cosmological dark matter. It links new symmetries of micro world to their dark matter signatures. 

Symmetry breaking induces new fundamental physical scales in
particle theory.
If the symmetry is spontaneously broken, it is restored, when the temperature exceeds the corresponding scale. In the course of cosmological expansion the temperature decreased and the transition to the phase with broken symmetry took place. depending on the symmetry breaking pattern, to formation of
topological defects in very early Universe. Defects can represent
new forms of stable particles (as it is in the case of magnetic
monopoles 
\cite{t'Hooft,polyakov,kz}), 
or extended macroscopic structures as cosmic strings \cite{zv1,zv2} 
or cosmic
walls \cite{okun}. Even unstable defects can leave replica in primordial nonlinear structures that remain in the Universe after the structure of defects decay (see below Sec.\ref{axion} and Ref.\citen{symmetry1} for recent review). Here we give a brief review of various forms of dark matter reflections of particle symmetry.
\section{Stable particles \label{pattern}}
Most of the known particles are unstable. For a particle with the
mass $m$ the particle physics time scale is $t \sim 1/m$
\footnote{Here and further, if it isn't specified otherwise we use the units $\hbar=c=k=1$}, so in
particle world we refer to particles with lifetime $\tau \gg 1/m$
as to metastable. To be of cosmological significance in the Big Bang Universe metastable
particle should survive after $t \sim (m_{Pl}/m^2)$, when the temperature of the Universe $T$
fell down below $T \sim m$ and particles go out of thermal equilibrium. It means that the particle lifetime
should exceed $t \sim (m_{Pl}/m) \cdot (1/m)$ and such a long
%
%
lifetime should be explained by the existence of a
symmetry. From this viewpoint, physics of dark matter is sensitive to the conservation laws
reflecting strict or nearly strict symmetries of particle theory.

\subsection{Weakly interacting massive particle miracle}
 \label{WIMPs}
The simplest form of dark matter candidates is the gas of new
stable neutral massive particles, originated from early Universe. Their stability can be protected by some discrete (as R-parity in supersymmetry) or continuous symmetry.

For
particles with the mass $m$, at high temperature $T>m$ the
equilibrium condition, $n \cdot \sigma v \cdot t > 1$ is valid, if
their annihilation cross section $\sigma > 1/(m m_{Pl})$ is
sufficiently large to establish the equilibrium. At $T<m$ such
particles go out of equilibrium and their relative concentration
freezes out. If particles have mass in the range of tens-hundreds GeV and annihilation cross section corresponding to weak interaction, the primordial frozen out abundance of such Weakly Interacting Massive Particles (WIMPs) can explain the observed dark matter density. This
is the main idea of the so called \textit{WIMP miracle}
 (see e.g. Refs.
\citen{book,newBook,DMRev,DDMRev} for details).

The process of WIMP annihilation to ordinary particles, considered in $t$-channel,
determines their scattering cross section on ordinary particles and thus
relates the primordial abundance of WIMPs to their scattering rate in the
ordinary matter. Forming nonluminous massive halo of our Galaxy, WIMPs can penetrate
the terrestrial matter and scatter on nuclei in underground detectors. The strategy of
direct WIMP searches implies detection of recoil nuclei from this scattering.

The process inverse to annihilation of WIMPs corresponds to their production in collisions
of ordinary particles. It should lead to effects of missing mass and energy-momentum,
being the challenge for experimental search for production of dark matter candidates at accelerators,
e.g. at LHC.

\subsection{Super-WIMPs}
The maximal
temperature, which is reached in inflationary Universe, is the
reheating temperature, $T_{r}$, after inflation. So, the very
weakly interacting particles with the annihilation cross section
$\sigma < 1/(T_{r} m_{Pl}),$ as well as very heavy particles with
the mass $m \gg T_{r}$ can not be in thermal equilibrium, and the
detailed mechanism of their production should be considered to
calculate their primordial abundance.

In particular, thermal production of gravitino in very early Universe is proportional to the reheating temperature $T_{r}$, what puts upper limit on this temperature from constraints on primordial gravitino abundance\cite{khlopovlinde,khlopovlinde2,khlopovlinde3,khlopov3,khlopov31,Karsten,Kawasaki}.
\section{Global U(1) symmetry \label{axion}}
A wide class of particle models possesses a symmetry breaking
pattern, which can be effectively described by
pseudo-Nambu--Goldstone (PNG) field (see Refs. \citen{DMRev,book2,PBHrev} for review and references). The coherent oscillations of this field represent a specific type
of cold dark matter (CDM) in spite of a very small mass of PNG particles $m_a=\Lambda^2/f$, where $f \gg \Lambda$, since these particles are created in Bose-Einstein condensate in the ground state, i.e. they are initially created as nonrelativistic in the very early Universe.
This feature, typical for invisible axion models can be the general feature for all the axion-like PNG particles.

At high temperatures the pattern of successive spontaneous and manifest breaking of global U(1) symmetry implies the
succession of second order phase transitions. In the first
transition at $T \sim f$, continuous degeneracy of vacua leads, at scales
exceeding the correlation length, to the formation of topological
defects in the form of a string network; in the second phase
transition at $T \sim \Lambda \ll f$, continuous transitions in space between degenerated
vacua form surfaces: domain walls surrounded by strings. This last
structure is unstable, but, as was shown in the example of the
invisible axion \cite{Sakharov2,kss,kss2}, it is reflected in the
large scale inhomogeneity of distribution of energy density of
coherent PNG (axion) field oscillations. This energy density is
proportional to the initial value of phase, which acquires dynamical
meaning of amplitude of axion field, when axion mass $m_a=C m_{\pi}f_{\pi}/f$ (where $m_{\pi}$ and $f_{\pi}\approx m_{\pi}$ are the pion mass and constant, respectively, the constant $C\sim 1$ depends on the choice of the axion model and $f\gg f_{\pi}$ is the scale of the Peccei-Quinn symmetry breaking) is switched on
in the result of the second phase transition.

The value of phase changes by $2 \pi$ around string. This strong
nonhomogeneity of phase leads to corresponding nonhomogeneity of
energy density of coherent PNG (axion) field oscillations. Usual
argument (see e.g. Ref. \citen{kim} and references therein) is essential
only on scales, corresponding to mean distance between strings.
This distance is small, being of the order of the scale of
cosmological horizon in the period, when PNG field oscillations
start. However, since the nonhomogeneity of phase follows the
pattern of axion string network this argument misses large scale
correlations in the distribution of oscillations' energy density.

Indeed, numerical analysis of string network (see review in the
Ref. \citen{vs}) indicates that large string loops are strongly suppressed
and the fraction of about 80\% of string length, corresponding to
long loops, remains virtually the same in all large scales. This
property is the other side of the well known scale invariant
character of string network. Therefore the correlations of energy
density should persist on large scales, as it was revealed in Refs.
\citen{Sakharov2,kss,kss2}. Discussion of such primordial inhomogeneous structures of dark matter
go beyond the scope of the present paper and we can recommend the interested reader
Refs. \citen{DMRev,book2,PBHrev} for review and references.
\section{New gauge symmetries}\label{mirror}
Extensive hidden sector of particle theory can provide the existence of new interactions, which only new particles possess. Historically one of the first examples of such self-interacting dark matter was presented by the model of mirror matter. Mirror particles, first proposed in Ref. \citen{LeeYang} to restore equivalence of left- and right-handed co-ordinate systems, represent a new set of symmetric partners for ordinary quarks and leptons\cite{KOP} with their own strong, electromagnetic and weak mirror interactions. It means that there should exist mirror quarks, bound in mirror nucleons by mirror QCD forces and mirror atoms, in which mirror nuclei are bound with mirror electrons by mirror electromagnetic interaction \cite{ZKrev,FootVolkas}. If gravity is the only common interaction for ordinary and mirror particles, mirror matter can be present in the Universe in the form of elusive mirror objects, having symmetric properties with ordinary astronomical objects (gas, plasma, stars, planets...), but causing only gravitational effects on the ordinary matter \cite{Blin1,Blin2}.

Even in the absence of any other common interaction except for gravity, the observational data on primordial helium abundance and upper limits on the local dark matter seem to exclude mirror matter, evolving in the Universe in a fully symmetric way in parallel with the ordinary baryonic matter\cite{Carlson,FootVolkasBBN}. The symmetry in cosmological evolution of mirror matter can be broken either by initial conditions\cite{zurabCV,zurab}, or by breaking mirror symmetry in the sets of particles and their interactions as it takes place in the shadow world\cite{shadow,shadow2}, arising in the heterotic string model. We refer to Refs.
\citen{newBook,OkunRev,Paolo} for current review of mirror matter and its cosmology.

If new particles possess new $y$-charge, interacting with massless bosons or intermediate bosons with sufficiently small mass ($y$-interaction),  for slow $y$-charged particles Coulomb-like factor
of "Gamov-Sommerfeld-Sakharov enhancement" \cite{Som,Sak,Sakhenhance} should be added in
the annihilation cross section
$$C_y=\frac{2 \pi \alpha_y/v}{1 - \exp{(-2 \pi \alpha_y/v)}},$$
where $v$ is relative velocity and $\alpha_y$ is the running gauge constant of $y$-interaction. This factor may not be essential in the period of particle freezing out in the early Universe (when $v$ was only few times smaller than $c$), but can cause strong enhancement in the effect of annihilation of nonrelativistic dark matter particles in the Galaxy. Products of annihilation contribute fluxes of cosmic rays and/or cosmic gamma radiation, giving a sensitive probe for even subdominant dark matter component \cite{ZKKC,DKKM}.  

\section{Approximate symmetries}

\subsection{Decaying dark matter}
Decaying particles with lifetime $\tau$, exceeding the age of the
Universe, $t_{U}$, $\tau > t_{U}$, can be treated as stable. By
definition, primordial stable particles survive to the present time
and should be present in the modern Universe. The net effect of
their existence is given by their contribution into the total
cosmological density. However, even small effect of their decay
can lead to significant contribution to cosmic rays and gamma background\cite{ddm}.
Leptonic decays of dark matter are considered as possible explanation of
the cosmic positron excess, measured in the range above 10 GeV by PAMELA\cite{pamela}, FERMI/LAT\cite{lat} and AMS02\cite{ams2}.

Primordial unstable particles with the lifetime, less than the age
of the Universe, $\tau < t_{U}$, can not survive to the present
time. But, if their lifetime is sufficiently large to satisfy the
condition $\tau \gg (m_{Pl}/m) \cdot (1/m)$, their existence in
early Universe can lead to direct or indirect traces\cite{khlopov7}.

Weakly interacting particles, decaying to invisible modes, can influence Large Scale Structure formation.
Such decays prevent formation of the structure, if they take place before the structure is formed.
Invisible products of decays after the structure is formed should contribute in the cosmological dark energy.
The Unstable Dark matter scenarios\cite{Sakharov1,UDM,UDM1,UDM2,UDM3,berezhiani4,berezhiani5,TSK,GSV} implied weakly interacting particles that form the structure on the matter dominated stage and then decay to invisible modes after the structure is formed.

Cosmological
flux of decay products contributing into the cosmic and gamma ray
backgrounds represents the direct trace of unstable particles\cite{khlopov7,sedelnikov}. If
the decay products do not survive to the present time their
interaction with matter and radiation can cause indirect trace in
the light element abundance\cite{khlopovlinde3,khlopov3,khlopov31,DES} or in the fluctuations of thermal
radiation\cite{UDM4}.
\subsection{Charge asymmetry of dark matter}
The fact that particles are not absolutely stable means that the corresponding charge is not strictly conserved and generation particle charge asymmetry is possible, as it is assumed for ordinary baryonic matter. At sufficiently strong particle annihilation cross section excessive particles (antiparticles) can dominate in the relic density, leaving exponentially small admixture of their antiparticles (particles) in the same way as primordial excessive baryons dominate over antibaryons in baryon asymmetric Universe. In this case {\it Asymmetric dark matter} doesn't lead to significant effect of particle annihilation in the modern Universe and can be searched for either directly in underground detectors or indirectly by effects of decay or condensation and structural transformations of e.g. neutron stars (see Ref. \citen{adm} for recent review and references). If particle annihilation isn't strong enough, primordial pairs of particles and antiparticles dominate over excessive particles (or antiparticles) and this case has no principle difference from the charge symmetric case. In particular, for very heavy charged leptons (with the mass above 1 TeV), like "tera electrons"\cite{Glashow}, discussed in \ref{asymmetry}, their annihilation due to electromagnetic interaction is too weak to provide effective suppression of primordial tera electron-positron pairs relative to primordial asymmetric excess\cite{BKSR1}.
\section{Dark atoms}
New particles with electric charge and/or strong interaction can
form anomalous atoms and contain in the ordinary matter as anomalous
isotopes. For example, if the lightest quark of 4th generation (that possess new conserved charge) is
stable, it can form stable charged hadrons, serving as nuclei of
anomalous atoms of e.g. anomalous helium
\cite{BKSR1,BKS,BKSR,FKS,I,BKSR4}. Therefore, stringent upper limits on anomalous isotopes, especially, on anomalous hydrogen put severe constraints on the existence of new stable charged particles. However, as we discuss in the rest of this review, stable doubly charged particles can not only exist, but even dominate in the cosmological dark matter, being effectively hidden in neutral "dark atoms"\cite{DADM}.

\subsection{Charged constituents of Dark Atoms}\label{asymmetry}
New stable particles may possess new U(1)
gauge charges and bind by Coulomb-like forces in composite dark
matter species. Such dark atoms cannot be luminous, since they
radiate invisible light of U(1) photons. Historically mirror matter
(see subsubsection \ref{mirror} and Refs. \citen{book,OkunRev} for review and references) seems to be the
first example of such an atomic dark matter.

However, it turned out that the possibility of new stable electrically charged leptons and quarks is not completely excluded and Glashow's tera-helium\cite{Glashow} has offered a new solution for this type of
dark atoms of dark matter. Tera-$U$-quarks with electric charge +2/3
formed stable ($UUU$) +2 charged "clusters" that formed with two -1
charged tera-electrons E neutral [($UUU$)$EE$] tera-helium "atoms" that
behaved like Weakly Interacting Massive Particles (WIMPs). The main
problem for this solution was to suppress the abundance of
positively charged species bound with ordinary electrons, which
behave as anomalous isotopes of hydrogen or helium. This problem
turned to be unresolvable\cite{BKSR1}, since the model\cite{Glashow}
predicted stable tera-electrons $E^-$ with charge -1.
As soon as primordial helium is formed in the Standard Big Bang
Nucleosynthesis (SBBN) it captures all the free $E^-$ in positively
charged $(He E)^+$ ion, preventing any further suppression of
positively charged species. Therefore, in order to avoid anomalous
isotopes overproduction, stable particles with charge -1 (and
corresponding antiparticles) should be absent, so that stable
negatively charged particles should have charge -2 only.

Elementary particle frames for heavy stable -2 charged species are
provided by: (a) stable "antibaryons" $\bar U \bar U \bar U$ formed
by anti-$U$ quark of fourth generation\cite{Q,I,BKSR4,Belotsky:2008se,DADM}
(b) AC-leptons\cite{DADM,FKS}, predicted in the
extension \cite{FKS} of standard model, based on the approach of
almost-commutative geometry\cite{bookAC}.  (c) Technileptons and
anti-technibaryons \cite{KK} in the framework of walking technicolor
models (WTC)\cite{Sannino:2004qp,Hong:2004td,Dietrich:2005jn,Dietrich:2005wk,Gudnason:2006ug,Gudnason:2006yj}. (d) Finally, stable charged
clusters $\bar u_5 \bar u_5 \bar u_5$ of (anti)quarks $\bar u_5$ of
5th family can follow from the approach, unifying spins and charges\cite{Norma}. Since all these models also predict corresponding +2
charge antiparticles, cosmological scenario should provide mechanism
of their suppression, what can naturally take place in the
asymmetric case, corresponding to excess of -2 charge species,
$O^{--}$. Then their positively charged antiparticles can
effectively annihilate in the early Universe.

If new stable species belong to non-trivial representations of
electroweak SU(2) group, sphaleron transitions at high temperatures
can provide the relationship between baryon asymmetry and excess of
-2 charge stable species, as it was demonstrated in the case of WTC
in Refs. \citen{KK,Levels1,KK2,unesco,iwara,I2}.

\section{Multicomponent Dark Matter}
Higher symmetry extensions of SM can embed various forms of dark matter candidates in a unique theoretical framework. 

Broken $SU(3)_H$ family symmetry not only described
the existence and observed properties of the three known quark-lepton families \cite{berezhiani4,berezhiani5,Berezhiani1,Berezhiani2} (see also \cite{bai1,bai2,bai3}), but also provided the physical mechanisms for inflation and baryosynthesis as well as it offered unified description of axion and massive neutrinos - candidates for Cold, Warm, Hot and Unstable Dark Matter. The parameters of axion cold dark matter (CDM), as well as the masses and lifetimes of neutrinos corresponded to the hierarchy of breaking of the $SU(3)_H$ symmetry of families, fixing their relative contribution into the total density. This approach gave a flavor of a quantitatively definite multi-component dark matter scenarios and elaborated the method to treat such multi-parameter models in an over-determined set of their physical, astrophysical and cosmological probes. It was considered as a bottom-up approach to heterotic string phenomenology, in which all the richness of of possible dark matter candidates can find their proper place \cite{Sakharov1}. 

Indeed $E_8$x$E'_8$ model combines supersymmetric candidates, 248 gauge bosons of $E'_8$ new interactions together with the set of 248 fundamental particles of shadow world. Embedding SM symmetry it can also contain additional quark-lepton family with its new gauge $U(1)$ interaction \cite{Kogan1,Shibaev}. Compactification of extra dimensions can lead to existence of homotopically stable objects \cite{Kogan2}. Multiple Kaluza-Klein (KK) dark matter candidates arise naturally in generic Type-IIB string theory compactification scenarios \cite{anupam}.Treatment of such multi-parameter space needs special methods developed by cosmoparticle physics.

\section{Towards cosmoparticle physics of dark matter\label{Discussion}}

The widely discussed complementarity of direct and indirect dark matter searches represents the simplest example of general methods of cosmoparticle physics, studying the fundamental basis and mutual
relationship between micro-and macro-worlds in the proper
combination of physical, astrophysical and cosmological signatures \cite{ADS,MKH,book,newBook}. Methods of cosmoparticle physics confronting the multi-parameter space of new phenomena, predicted by particle theory, with the over-determined set of their physical, astrophysical and cosmological probes can give in their development clear answer on the true picture of the Universe and physical laws, on which it is based.
In particular, it will shed light on the problem of cosmological dark matter in the context of the fundamental structure of the microworld.

\section*{Acknowledgments}

The work was performed within the framework of the Center FRPP supported
by MEPhI Academic Excellence Project (contract 02.03.21.0005,
27.08.2013).
%


\end{document}